\documentclass[%
 reprint,
 twocolumn,
 amsmath,amssymb,
 aps,
 prl,
 nofootinbib,
 tightenlines,
 nobibnotes
]{revtex4-2}

\usepackage{graphicx}
\usepackage{dcolumn}
\usepackage{bm}
\usepackage{hyperref}

\usepackage{color}

\usepackage{gensymb}

\definecolor{Green}{RGB}{0,125,0}
\definecolor{Blue}{RGB}{0,0,255}
\definecolor{LBlue}{RGB}{150,200,255}
\definecolor{Red}{RGB}{200,010,010}
\definecolor{DGreen}{RGB}{10,150,10}
\definecolor{DarkRed}{RGB}{251,0,250}
\newcommand{\ms}[1]{{\textcolor{Blue}{#1}}}

\begin{document}

\title{Measurement of nonequilibrium vortex propagation dynamics in a nonlinear medium}

\author{Patrick C. Ford$^1$}
\author{Andrew A. Voitiv$^1$}
\author{Chuanzhou Zhu$^2$}
\author{Mark T. Lusk$^2$}
\author{Mark E. Siemens$^1$}
\email[]{msiemens@du.edu}
\affiliation{$^1$Department of Physics and Astronomy, University of Denver, \\ 2112 E. Wesley Avenue, Denver, CO 80208, USA
}%
\affiliation{
$^2$Department of Physics, Colorado School of Mines, \\ 1500 Illinois Street, Golden, CO 80401, USA
}%

\date{\today}

\begin{abstract}
We observe and measure the nonequilibrium dynamics of optical vortices as a function of propagation distance through a nonlinear medium. The precession of a tilted-core vortex is quantified as is vortex-core sharpening, where the infinite width of a linear core subsequently shrinks and approaches the healing length of this nonlinear optical fluid. Experiments are performed with a variable-length nonlinear medium: a nonlinear fluid
in a tank with an output window on a translating tube.
This provides control over the distance the light propagates in the fluid and allows for the measurement of the dynamics throughout the entire propagation range. Results are compared to the predictions of a computational simulator to find the equivalent dimensionless nonlinear coefficient.
\end{abstract}

\maketitle


\section{\label{sec:1}Introduction}

In quantum fluids such as Bose-Einstein condensates (BEC) or superfluid helium, vortex cores can be characterized by a size $\ell$ that is usually taken to be the healing length $\xi$. In those systems, dynamical measurements observe vortex cores with constant $\ell$ and ellipticity because vortices are generated by stirring or other non-deterministic means \cite{madison2000vortex,matthews1999vortice,eckel2014hysteresis,white2010nonclassical} and quickly evolve to the equilibrium core size $\xi$ before measurements can be made. Programmatic control over the core shape of initial vortex states with the ability to directly measure propagation dynamics would enable the observation of the formation of vortex solitons, and provide insight into the nucleation and annihilation processes of these vortex solitons.

The propagation of structured light through nonlinear media offers a particularly useful setting for studying the mean-field behavior of quantum fluids. A wide range of initial conditions can be easily programmed using a holographic transfer function, enabling arbitrary control over initial vortex amplitude profiles. In addition, the propagation axis amounts to a pseudo-time, so that any given time slice can be interrogated in detail. In linear optics, studies have been performed both numerically \cite{Rozas1997} and experimentally \cite{AndersenJasmineM2023Asoo} to investigate how the size and tilt of  small-core optical vorticies evolve as they propagate. These cores expand with propagation, approaching the infinite width of a linear core, a fundamental mode of linear propagation. 

\begin{figure}[h!]
\centering
\includegraphics[width=\linewidth]{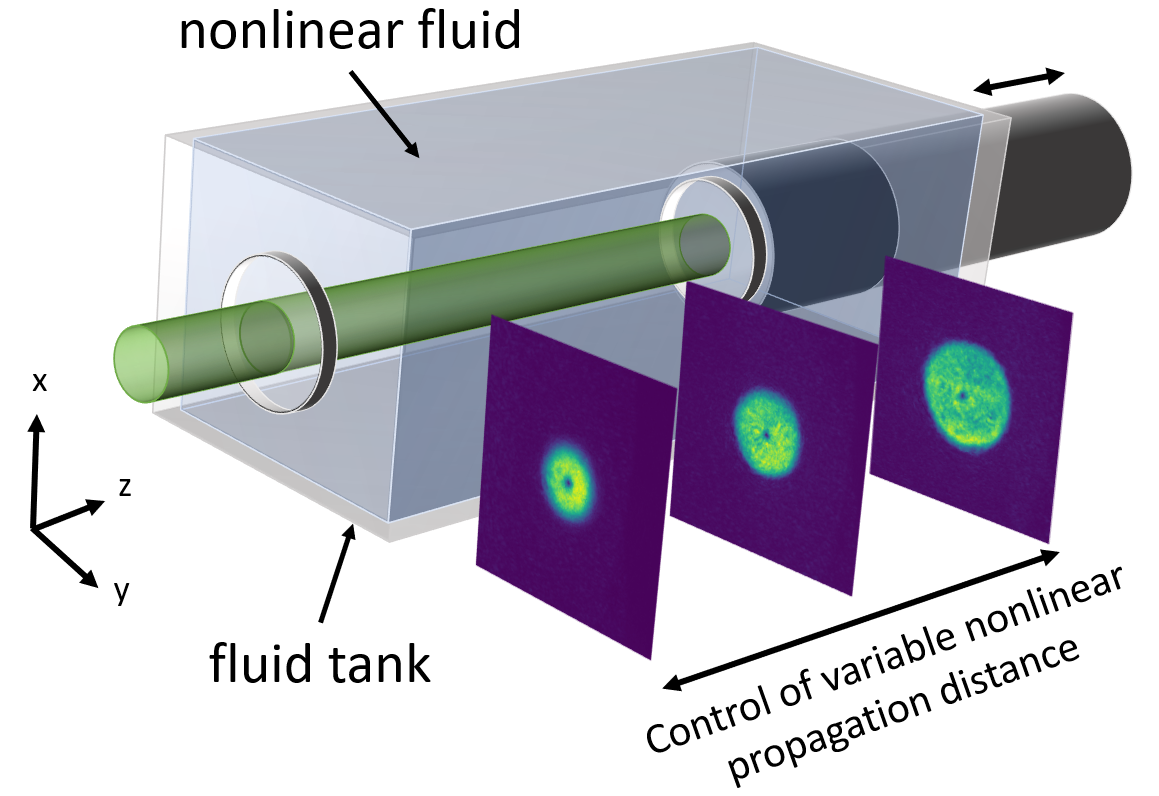}
\caption{
Variable-propagation nonlinear cell. A tube equipped with an output window can be translated within a $\chi^{(3)}$ fluid to observe vortex structure at any point along the pseudo-time axis. Example intensity profiles are shown for minimum, medium, and maximum nonlinear propagation distances. Medium propagation distance corresponds to a minimum core width. 
}
\label{fig:2}
\end{figure}

There is significant interest in extending these sorts of measurements to \emph{nonlinear} optics, where interactions enable a formal analog between propagating light and quantum fluid dynamics\cite{CarusottoIacopo2013Qfol}. Nonlinear optical vortex experiments have been conducted using fixed physical propagation distances in gaseous\cite{bakerrasooli2023turbulent,AzamPierre2022Vcaa}, liquid\cite{Rozas:00,VockeDavid2016Rogi}, and solid media\cite{Engay:20}. Nonlinear propagation studies have been approximated by pseudo-imaging various planes through the medium \cite{VockeDavid2015Econ}, or by varying the strength of the nonlinearity as a surrogate to varying the propagation distance\cite{FontaineQ2018OotB}.  In the former, a portion of the imaging system path must lie within the nonlinear medium, resulting in recovery of approximations of the object plane. In the latter, the strength of the nonlinearity changes fundamental properties of the nonlinear dynamics, such as the healing length that characterizes core-width. To the best of our knowledge, the propagation dynamics of an initially infinite-width core propagating through a nonlinear medium have yet to be directly observed. In a defocusing Kerr-type medium, such an observation would be particularly interesting as it is the formation of a dark soliton\cite{DesyatnikovAntonS.2005Ovav}, analogous to the dark solitons of two-dimensional BECs which exhibit particle-like interactions\cite{Zhu_2022}.

In this paper, we directly measure propagation dynamics of single-vortex laser beams in a nonlinear Kerr-type medium. These measurements are made possible by a novel apparatus using a fluid medium of variable length in the optical axis direction, as shown in Fig. \ref{fig:2}. This novel apparatus is used to quantify the evolution of an optical vortex that  initially has a linear amplitude profile. We find that its propagation through a medium with a negative second-order refractive index exhibits core-sharpening. The precession of an initially tilted vortex under similar nonlinear propagation is also quantified.

Vortex structure is characterized by a phase singularity, surrounded by a phase gradient, along with an amplitude profile. For a circular vortex, the amplitude profile can generally be described as a monotonically increasing function of distance from the core singularity. For a single-charge vortex, as used in the experiments described here, linear media can only support equilibrium core amplitudes that increase linearly with distance from the singularity. For incompressible superfluid helium, this amplitude reaches the background density within a few Angstroms and is often approximated with an inverse delta function\cite{BridgesThomasJ.2017Apoq}. Equilibrium core profiles within compressible Bose-Einstein condensates and nonlinear photonic fluids are finite, having a ratio of system size $R$ to healing length $\xi$ on the order $R/\xi \sim 10-100$. Vortices in such continua therefore lie between the infinite width of the linear core and the infinitely small point core of superfluid helium. The finite size of the core width depends on the strength of the nonlinearity\cite{SoninEdouardB.2015DoQV}.

A formal analogy can be drawn between nonlinear optical propagation described by the Nonlinear Schrodinger Equation (NLSE)\cite{CarusottoIacopo2014Slib} and the time evolution of the mean field of a quantum fluid described by the two-dimensional Gross-Pitaevskii Equation (GPE)\cite{GrossE.P.1961Soaq,DzyaloshinskiiIE1961GTOV}. The GPE predicts that the amplitude profile surrounding the phase singularity of a wide-core vortex will become steeper with propagation, narrowing the core width and approaching the healing length of the fluid \cite{BridgesThomasJ.2017Apoq}. In addition to such core sharpening, the model predicts that a tilted core will precess as it propagates \cite{ZhuChuanzhou2021Doev}. A correspondence is constructed that allows the non-dimensional GPE coefficients to be expressed in terms of physical parameters, facilitating direct, quantitative comparison and fitting of optical measurements to model predictions.

\section{\label{sec:2}Theory for nonlinear optical propagation and mean-field quantum fluid dynamics}
The paraxial propagation of a classical laser beam, having a spatial description of the electric field $E(\textbf{r},z)$ in a nonlinear $\chi^{(3)}$ medium, is described by the NLSE \cite{CarusottoIacopo2014Slib};
\begin{equation}\label{eq:1}
    i\partial_z E = \bigg(-\frac{1}{2k_0n_0}\nabla_\perp^2 + V(\textbf{r},z) - \frac{k_0}{2n_0^2}\chi^{(3)}|E|^2\bigg)E.
\end{equation}

 Here the parameters $x$, $y$ and $z$ are distances along orthogonal spatial directions with $z$ being the axis of optical propagation. Lateral position within a given pseudo-time slice is thus described with $\textbf{r} = x\hat{x} + y\hat{y}$. The parameter $k_0 = 2\pi/\lambda_0$ is the wavenumber, with $\lambda_0$ the wavelength of the coherent light beam in free space. In the optical setting, the potential function  $V(\textbf{r},z)$ characterizes dielectric inhomogeneities that can be used to produce a trapping effect\cite{CarusottoIacopo2014Slib}. The parameters $n_0$ and $\chi^{(3)}$ 
 are the linear refractive index and the third-order susceptibility, respectively, with the latter accounting for self-interaction of the electric field.

Nondimensionalization of the NLSE can be carried out by choosing appropriate units for observable quantities. We outline one method of nondimensionalization in the Supplemental Material (SM) section S.1 that results in the description of a two-dimensional, time-evolving, dimensionless wavefunction $\psi'(\textbf{r}^{'},t')$ :
\begin{equation}\label{eq:2}
    i  \partial_{t'} \psi^{'}  = \bigg( -\frac{1}{2}\nabla_{\perp}^{'2} - \frac{4}{3}k^2 \frac{n_2}{n_0^2}\gamma^2 I_0 |\psi'|^2\bigg)\psi'.
\end{equation}
Here, $\textbf{r}'$ and $t'$ are dimensionless space and time parameters with $\nabla^{'2}_\perp$ and $\partial_{t'}$ the corresponding nondimensionalized derivative operators. The coefficient $\gamma = \sqrt{\pi}w_0^2/2$ ensures normalization of $\psi'$, the dimensionless wavefunction, and the parameter $k=k_0 n_0$ is the wavenumber within the medium.

Direct comparison with the GPE allows the fluid nonlinearity to be captured in a single nonlinearity coefficient:
\begin{equation}\label{eq:3}
    \beta = -\frac{4}{3}k_0^2n_2\gamma^2 I_0 .
\end{equation}
Our choice of nonlinear medium, described in Section 3, has a negative second-order index. The sign of our nonlinear term is therefore positive, corresponding to a repulsive interaction that supports the formation of solitons \cite{SWARTZLANDER1992}.

Numerical simulations were performed using a previously developed pseudo-spectral, two-dimensional GPE simulation tool\cite{AntoineXavier2014GaMt,AntoineXavier2015GaMt}. Consistent with the optical setting of Eq. \ref{eq:2}, the potential, $V$, was set to zero. The nondimensional grid size was $3.52\times10^{-2}$ in the x- and y-dimensions, with a pseudo-time (z) step size of $1\times10^{-4}$. The value of $\beta$ was varied between $\beta=0$ and $\beta=20$. Initial conditions were prepared as a single-charge linear core for the core-sharpening experiments as described in Eq. \ref{eq:4}
and as a combination of positive and negative single-charge linear cores for the tilt precession experiments as shown in Eq. \ref{eq:9}. 
In all numerically simulated cases, the scaling used (see SM) is such that the beam waist 
$w_0=1$.

\section{\label{sec:3}Experimental apparatus}
\begin{figure}[h!]
\centering
\includegraphics[width=\linewidth]{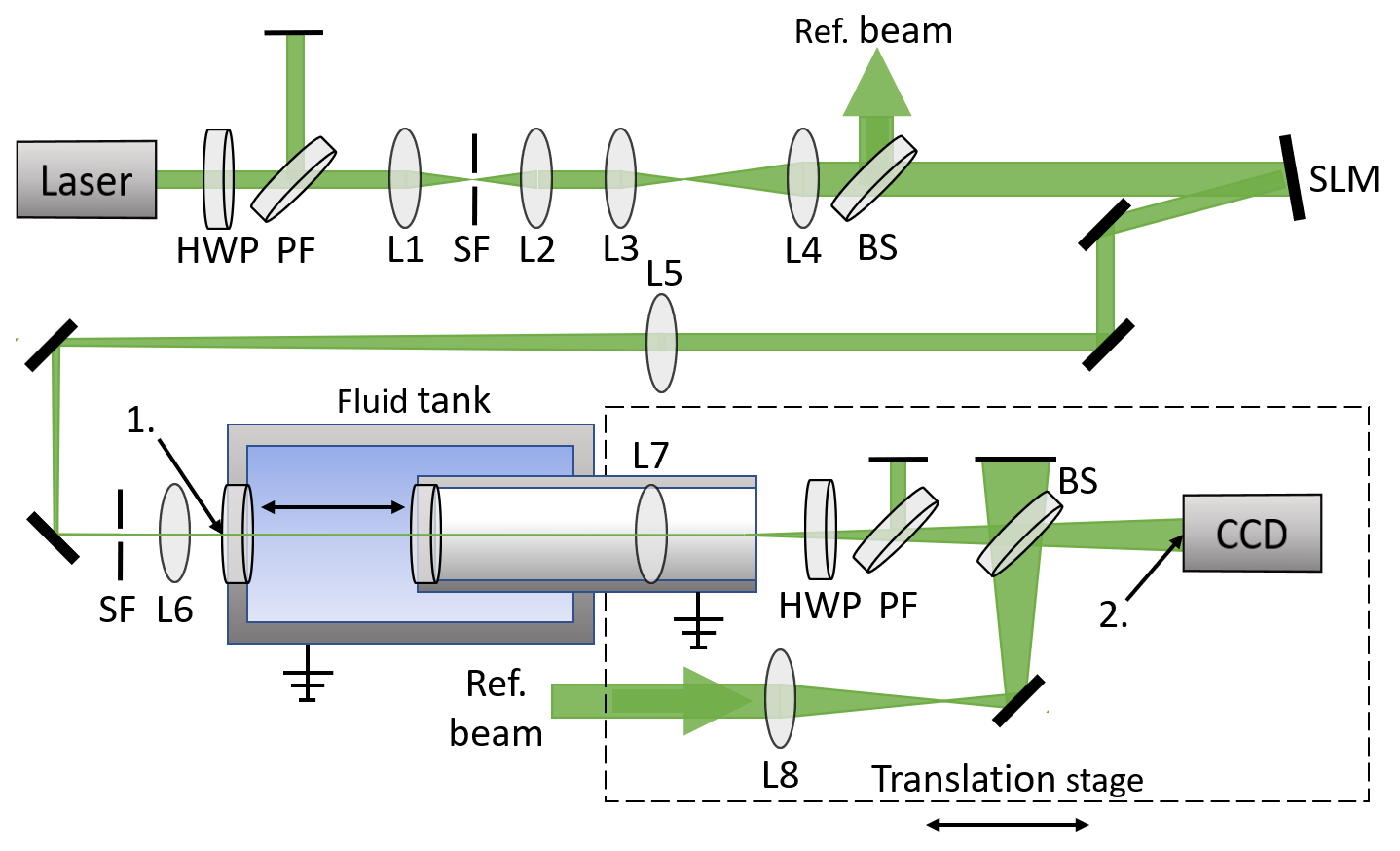}
\caption{
Schematic of the experimental apparatus. A high-power laser beam is attenuated by a half-wave plate (HWP) and polarization filter (PF), prepared by a lens pair (L1 and L2) and spatial filter (SF), and enlarged with a second lens pair (L3 and L4). A beam splitter (BS) picks off a reference beam for later interferometric phase imaging to recover the phase of the beam. The beam shines on a phase-only SLM, encoding the first diffracted order with the desired field. The first-diffracted order passes through a third lens pair (L5 and L6) where it is spatially selected, de-magnified ($M=0.1$), and imaged onto the input window of a nonlinear fluid tank. A translating imaging tube holds the output window and the lens of a single-lens imaging system (L7). The signal beam passes through an attenuation stage (HWP and PF) and a beam splitter (BS) where it is interfered with the reference beam and imaged onto a CCD camera sensor. The reference beam passes through a lens (L8) to match it's curvature to the signal beam. The SLM is imaged onto the input window at location 1. The output window is imaged onto the CCD sensor at location 2.}
\label{fig:1}
\end{figure}

A schematic diagram of the experimental apparatus is shown in Fig. \ref{fig:1}. A high-power ($15$ $W$) green ($532$ $nm$) laser is used as the source for these experiments. To control the power, the beam is passed through a rotatable waveplate and a fixed polarizing filter. To ensure an ideal Gaussian mode the beam is spatially filtered, recollimated and magnified ($\times$3) to fill a phase-only spatial light modulator (SLM) panel. The SLM  encodes the desired beam phase and amplitude profiles in the first diffracted order using a blazed grating \cite{MorenoIgnacio2020Deos} transfer function to maximize the power efficiency in the first diffracted order. A telescope spatially selects the first diffracted order and images the SLM panel onto the input window of the fluid tank. This telescope also demagnifies the image of the SLM ($\times0.1$) to increase the peak intensity of the signal, enhancing the strength of the nonlinearity by $100\times$.

The novel feature of our experimental apparatus is the use of a liquid medium and a tank designed to enable a variable path length through that medium. The tank is also modified to have an input window on one side. We use a custom-fabricated dynamic shaft seal and otherwise readily-available parts to allow a tube holding an output window to translate within the fluid tank, providing control over propagation distance. Additional details on the tank design are presented in the supplemental material, S.2.

The tube holding the output window contains a single-lens imaging system which magnifies ($\times 2.8$)
and images the output window onto a CCD camera sensor. The resolution limit ($2.7$ $\mu$m) of this imaging system is on the order of the CCD pixel size at the object plane ($2.3$ $\mu$m). A beam splitter is included in this imaging path to allow interference with the reference beam. The output imaging system includes its own waveplate attenuator consisting of a rotatable waveplate and fixed polarization filter, as well as neutral density filters to optimize interferogram contrast and camera exposure. The maximum practical beam waist at the input is $\sim$ $200$ $\mu$m while the minimum practical waist is $\sim$ $100$ $\mu$m. The maximum achieved characteristic beam intensity in this experiment was $\sim$ $2.5$ $\mathrm{kW/cm^2}$.

The telescoping tube and camera imaging system are mounted together on a motorized linear stage so that the camera sensor is always held in the image plane of the output window, for any tube z-scan position. This fixed imaging distance between output and camera enables direct measurement of the beam at the output window after an arbitrary propagation distance in the nonlinear medium. The minimum practical propagation distance is 17 mm, while the maximum is 117 mm.

The nonlinear fluid is a suspension of graphene oxide (GO) in ethanol. The graphene powder is prepared by grinding it with mortar and pestle and allowing the powder to oxidize freely in air. Approximately 1 g of GO powder is added to a volume of ethanol that fills the tank ($\sim 1.8$ L). The suspension is allowed to settle for a few minutes and is poured through a filter paper to remove the largest of the particles. After adding the supsension to the tank, a settling period of one hour is allowed, during which larger particles tend to fall out. Using Stoke's law to relate particle size to settling velocity, we estimate the size of particles remaining in suspension to be about 10 $\mu$m or smaller. Filtering and settling is performed to reduce experimental noise in the forms of 1.) shot-to-shot imaging measurements that can be affected by a large particle drifting across the beam, and 2.) scan-to-scan measurements, between which the large particles can settle out of suspension causing a change in the magnitude of the nonlinearity. Using Snell's law, the linear index of refraction of the GO suspension is measured to be $n_0 \approx 1.4$. Using the Beer-Lambert law, the absorption coefficient of the suspension is measured to be 1.02 m\textsuperscript{-1} giving a $1/e$ absorption length of 0.98 m. The absorption coefficient appears constant over the range of input intensities achievable by the experimental apparatus, and the $1/e$ absorption distance is about $10\times$ longer than the maximum propagation distance available in the tank. With these considerations, we assumed absorption to be linear and negligible, and excluded the absorption process in the GPE model for numerical simulations.

\begin{figure}[h!]
\centering
\includegraphics[width=\linewidth]{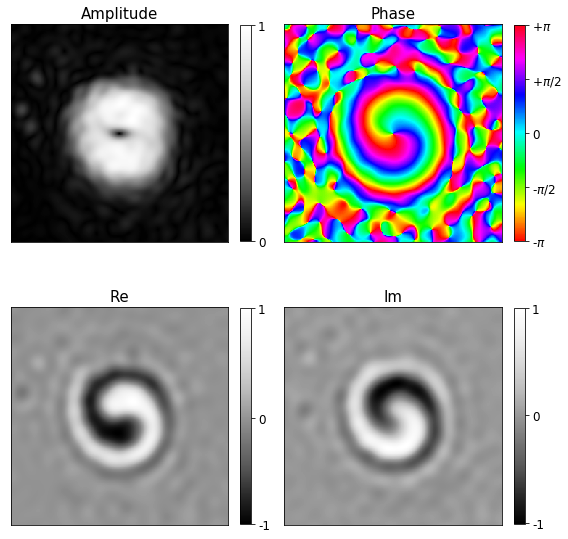}
\caption{
Example reconstruction of complex field captured after propagation in nonlinear medium. Input beam contains a single tilted core of Eq. \ref{eq:9} with a polar tilt angle $\theta=60^\circ$. The input beam beam waist and vortex azimuthal angle were set to $w_0=200$ $\mu$m and $\xi=0^\circ$ respectively. An interferogram was created using a reference beam and interference on a beam splitter. The full complex field was reconstructed using off-axis digital holography. Shown from left to right and top to bottom are the measured amplitude, phase, real part, and imaginary part of the field.}
\label{fig:3}
\end{figure}

A reference beam is split from the prepared beam before the SLM and then interfered on a beam splitter in the camera's imaging stage with the output signal from the tank to allow the capture of interferograms. 
The interferogram is digitally processed using single-shot off-axis digital holography \cite{PicartPascal2015Ntid} if recovery of the full complex field of the propagated beam is desired as shown for an example tilted vortex in Fig.\ref{fig:3}.


\section{\label{sec:4}Measurement of vortex core sharpening dynamics}
As a first demonstration of our setup for dynamic nonlinear propagation measurement, we injected a Laguerre-Gaussian\cite{ALLENL1992Oamo} (LG) mode at the tank input. An LG mode is characterized by having a phase vortex in a Gaussian background, or a background amplitude characterized by a Laguerre polynomial, with radial index $p$ and azimuthal index $m$: 
\begin{eqnarray}\label{eq:3.5}
    \mathrm{LG}(\textbf{r})_{m,p}  = C_{m,p}^{\mathrm{LG}}\frac{1}{w_0}\bigg(\frac{r\sqrt{2}}{w_0}\bigg)^{|m|} \nonumber \\ \times e^{r^2/w_0^2}L_p^{|m|}\bigg(\frac{2r^2}{w_0^2}\bigg) e^{-im\phi}
\end{eqnarray}
with $\textbf{r} = x\hat{x} + y\hat{y}$ being the position vector, $r = \sqrt{x^2+y^2}$ being the distance from the origin, $w_0$ being the beam waist, $\phi = \arctan(y/x)$ being the azimuthal angle, and $C_{m,p}^{LG} = \sqrt{\frac{2p!}{\pi(p+|m|)!}}$ being a normalization coefficient.

The LG input shape with azimuthal index $m = \pm1$ and radial index $p = 0$ (LG\textsubscript{$\pm$1,0}) used for the core-sharpening experiment represents a single linear core vortex in a Gaussian background:
\begin{eqnarray}\label{eq:4}
    \psi_0(x,y,z=0) &=& \mathrm{LG}_{\pm1,0}(x,y) \nonumber \\ &=& \frac{2}{\sqrt{\pi}w_0^2}\exp\bigg[-\frac{(x^2+y^2)}{w_0^2}\bigg]\big(x\mp iy\big).
\end{eqnarray}
For these experiments, the input beam width was set to $w_0 = 200$ $\mu$m.

\begin{figure}[h!]
\centering
\includegraphics[width=\linewidth]{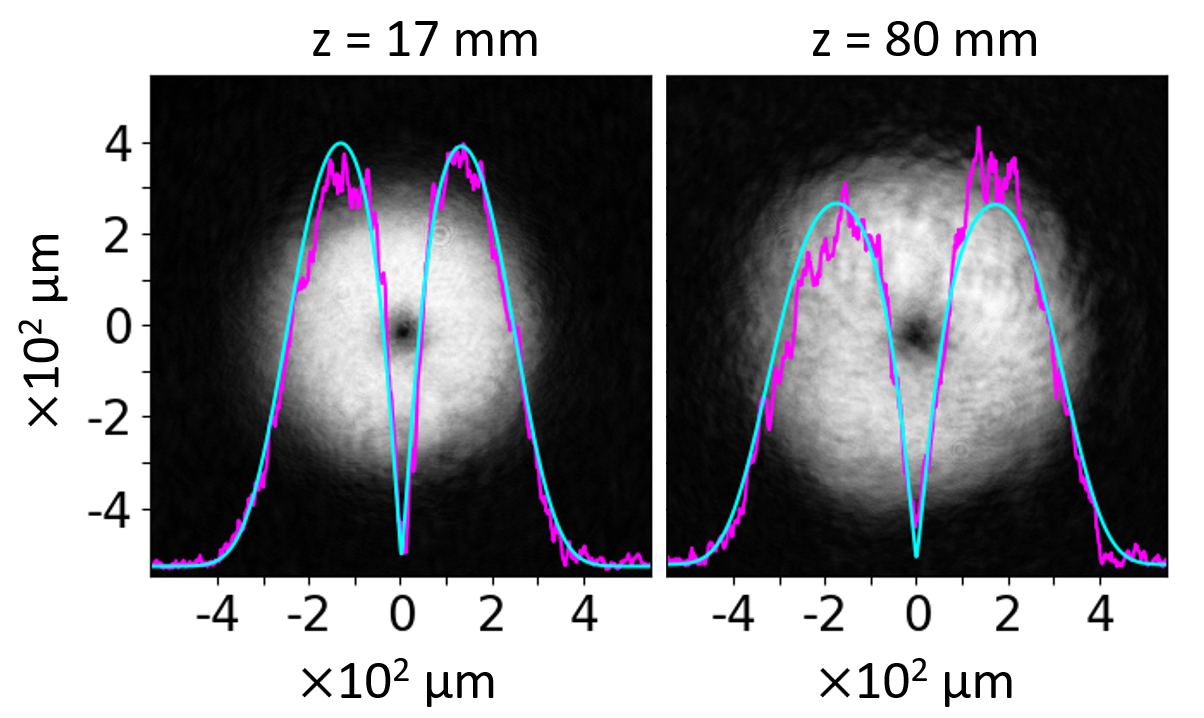}
\caption{Low-intensity (11.3 W/cm\textsuperscript{2}) beam evolution near the linear limit. Left panel shows the intensity profile of the beam at minimum propagation distance,  17 mm (0.05$z_R$). Right panel shows the intensity profile at 80 mm (0.25 $z_R$). The beam grows and evolves self-similarly as an eigenmode of linear propagation.}
\label{fig:4}
\end{figure}

\begin{figure}[h!]
\centering
\includegraphics[width=\linewidth]{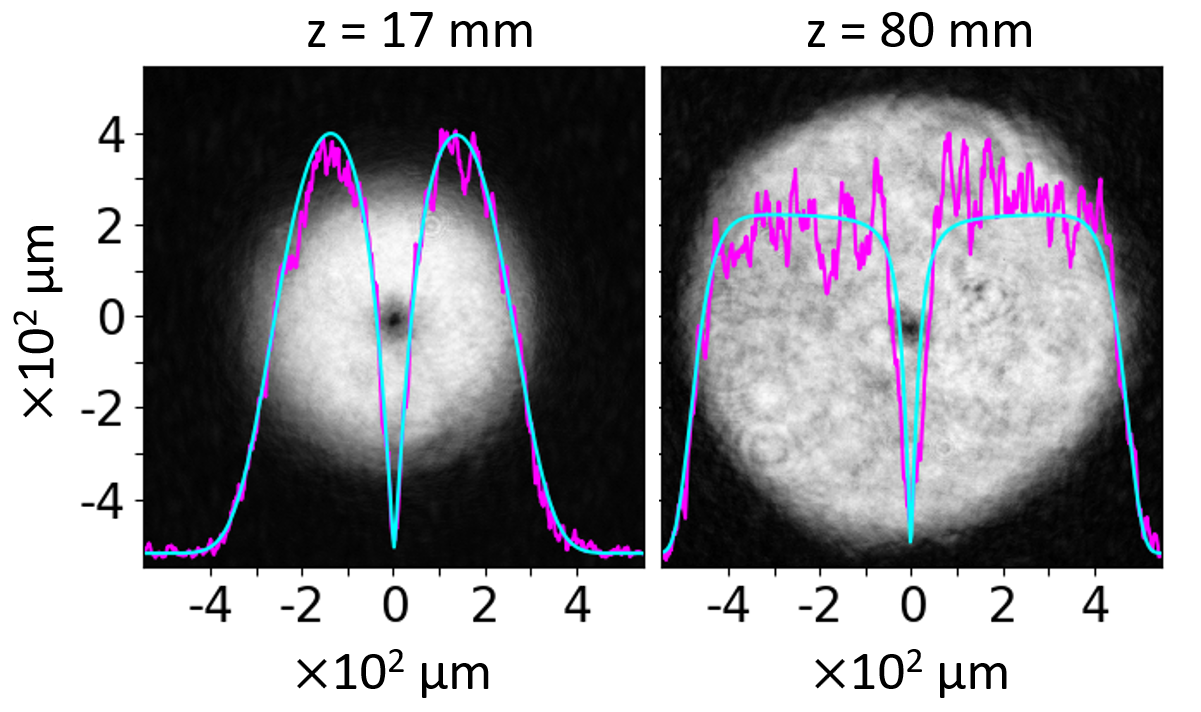}
\caption{High-intensity (2.2 kW/cm\textsuperscript{2}) beam evolution shows strongly nonlinear and nonequilibrium dynamics. Left panel shows the intensity profile of the beam at minimum propagation distance, 17 mm (0.05 $z_R$). Right panel shows the intensity profile at 80 mm (0.25 $z_R$), the propagation distance at which approximately the minimum core width occurred. The beam evolves according to the NLSE with $n_2<0$, with the core sharpening and the background expanding, becoming flatter near the center and steeper at the edges.}
\label{fig:5}
\end{figure}

While LG modes are eigenmodes of free space, the eigenmode inside the nonlinear medium should have a smaller core size. We expect 
and observe a dynamic sharpening of the core with nonlinear propagation of an initially LG beam. Some examples of beam evolution for linear and nonlinear propagation are shown in Fig. \ref{fig:4} and Fig. \ref{fig:5} respectively.

If we consider only the amplitude of the input LG\textsubscript{$\pm$1,0} modes, we have
\begin{equation}\label{eq:5}
    |\psi_0(r)| = \frac{2}{\sqrt{\pi}w_0^2}\exp\bigg[-\frac{r^2}{w_0^2}\bigg]r.
\end{equation}
Intensity profiles were measured at discrete z-steps within the medium from a minimum propagation distance of 17 mm to a maximum of 117 mm in steps of 2 mm. This corresponds to a range of 0.05 $z_R$ to 0.33 $z_R$, where $z_R = \pi w_0^2/\lambda$ is the linear diffraction length. We performed this experiment at three input characteristic intensities (11 W/cm\textsuperscript{2}, 71 W/cm\textsuperscript{2}, 2.2 kW/cm\textsuperscript{2}) to observe the effect of increasing nonlinearity.

As this beam propagates in the self-defocusing nonlinear medium, it will evolve such that the core amplitude gradient will steepen and the core will contract in size. To quantify the core width $\ell$, we fit the amplitude profile to that of a nonlinear core\cite{ZhuChuanzhou2021Doev},
\begin{equation}\label{eq:6}
    |\psi_v(r)| = N_v\sqrt{1 - \frac{\ell^2}{r^2+\ell^2}}.
\end{equation}
Here $N_v$ is a normalization coefficient for the core shape. As the background Gaussian of the input beam propagates in the self-defocusing nonlinear medium\ms{,} the amplitude gradient at the outer edges becomes steeper and the gradient near the center becomes flatter. We model this behaviour using the supergaussian
\begin{equation}\label{eq:7}
    |\psi_{bg}(r)| = N_{bg} \exp\bigg[-\bigg(\frac{r}{w}\bigg)^g\bigg],
\end{equation}
where $N_{bg}$ is a normalization coefficient for the background shape, $w$ is a propagated beam width, and the parameter $g$ allows the supergaussian to represent anything from a Gaussian ($g = 2$) to a flat-top beam ($g \xrightarrow{}\infty$). To fit the measured beam profiles, we used a composite nonlinear (LG\textsubscript{$\pm$ 1,0}) single-core amplitude shape given by
\begin{equation}\label{eq:8}
    |\psi(r)| = N\exp\bigg[-\bigg(\frac{r}{w}\bigg)^g\bigg]\sqrt{1 - \frac{\ell^2}{r^2+\ell^2}}.
\end{equation}

Images of the optical intensity were captured at 2 mm propagation steps in the nonlinear medium. Amplitude profiles were then digitally processed by cropping and converting from intensity to electric field amplitude by taking the square-root of the images. The electric field data was then fit to the shape of Eq. \ref{eq:8} to extract the core width $\ell$. The numerical model was also used to produce field solutions at stepped propagation distances and were fit to Eq. \ref{eq:8} using the same digital pipeline as the experimental data.

The shape of the core-width vs. propagation distance can be intuitively understood. With approximately linear propagation, the Laplacian term in the dynamics dominates and the core only grows. Under nonlinear propagation, the core width decreases as the nonlinear term in the dynamics dominates, approaching the healing length of the optical fluid \cite{BridgesThomasJ.2017Apoq}. With further unconfined propagation and nonlinear refraction, the peak beam intensity falls off, the Laplacian term begins to dominate over the nonlinear term, and the core width begins to grow. With a larger nonlinearity, the minimum core width is smaller.

\begin{figure}[h!]
\centering
\includegraphics[width=\linewidth]{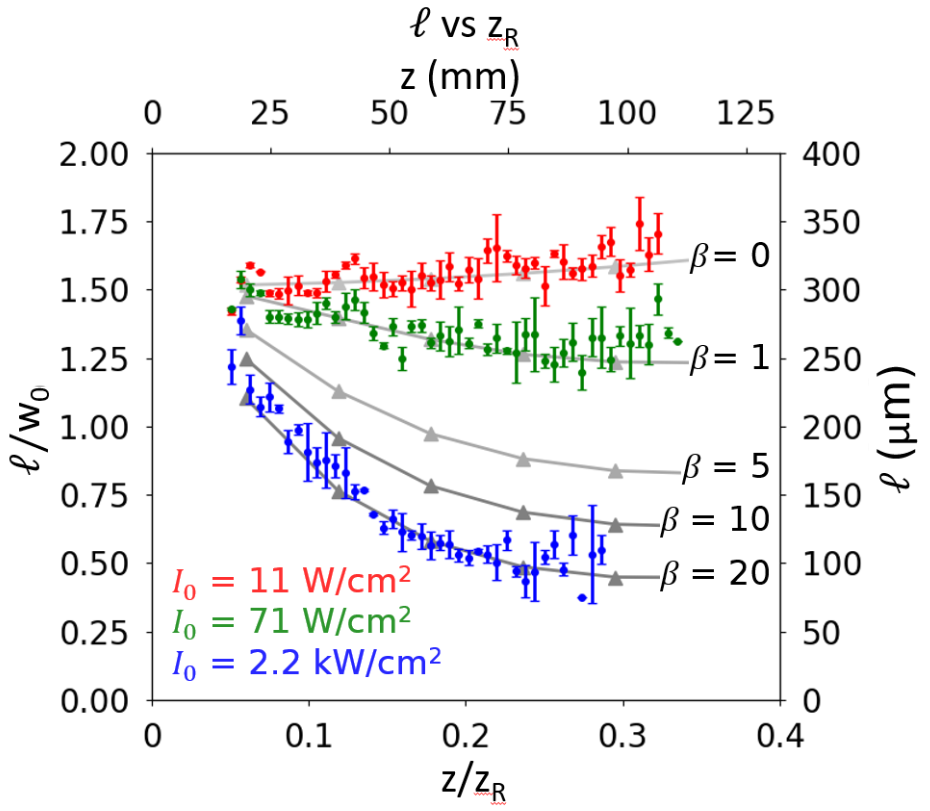}
\caption{Core-width $\ell$ vs propagation distance $z$ at various values of $\beta$ and input powers from simulation and experiment. Colored dots represent results from experiment, nondimensionalized by linear diffraction length $z_R = k w_0^2/2$ in the propagation direction and by the initial beam width $w_0$ in the transverse plane. Error bars represent one standard deviation over 5 trials. Gray triangles represent numerical results with connecting lines to guide the eye.}
\label{fig:6}
\end{figure}

The core widths extracted from the experimental data are plotted in Fig. \ref{fig:6} along with the core width 
obtained from the numerical simulation using matching initial conditions. For direct comparison, the experimental data and nondimensional numerical data are scaled by $z_R$ in the propagation direction and by $w_0$ in the transverse direction. At our minimum input intensity, the linear model matches our measurements of the core dynamics. As expected, the measurements with higher input power show significant core sharpening; the core sharpening dynamics match well with the model when nonlinearity is included, up to a conservative best-fit nonlinearity of $\beta=15$ for the 2.2 kW/cm\textsuperscript{2} input power, and a nonlinear index $n_2\approx -1.1 \times 10^{-9}$ cm\textsuperscript{2}/W. This estimate for $n_2$ agrees in magnitude with published values for GO suspensions of similar concentration \cite{KumarSunil2009Fcda,Jimenez2018Moon,Ebrahimi}.

\section{\label{sec:5}Measurement of Tilted-core precession}
\begin{figure}[h!]
\centering
\includegraphics[width=\linewidth]{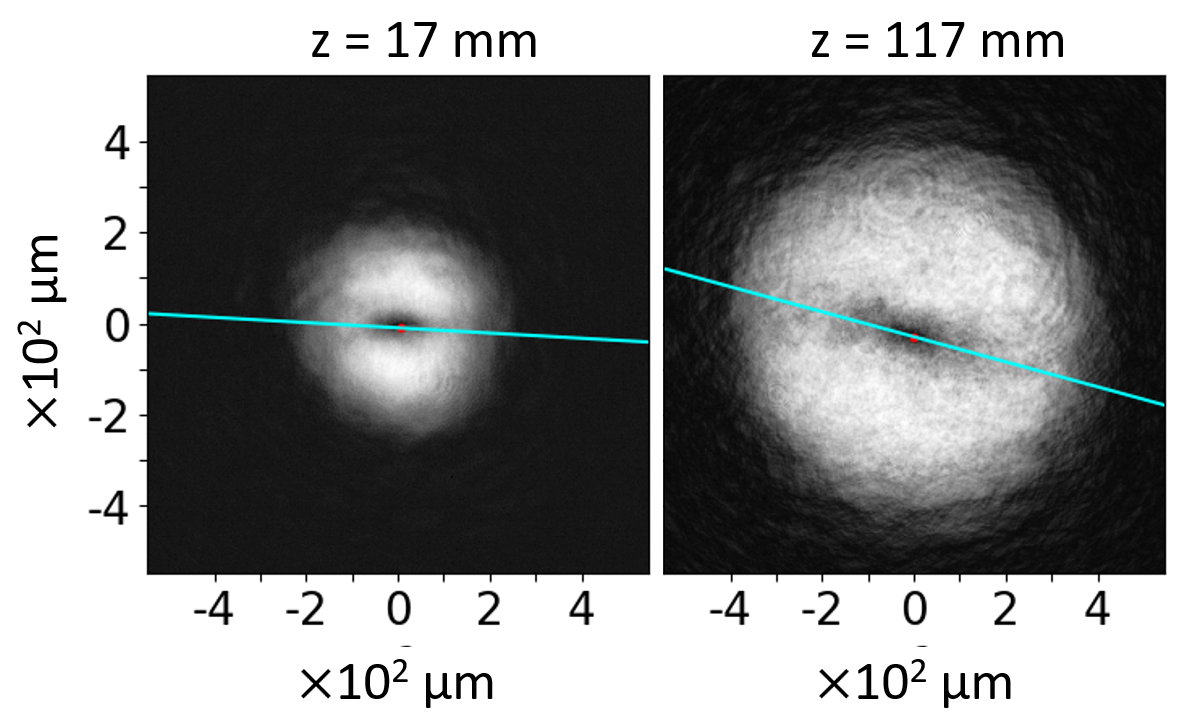}
\caption{Precession with propagation. Shown above are intensity profiles collected in the tilt-precession optical experiment. The left panel shows a short propagation distance, 17 mm, at high power. The right panel shows a long propagation distance, 117 mm, at high power. Cyan lines show the precession angle fit to the data. With propagation in the nonlinear Kerr-type medium, the tilted optical vortex precesses. The characteristic intensity of the input beam was approximately 2.5 kW/cm\textsuperscript{2} with a beam waist of $w_0=100$ $\mu$m.}
\label{fig:7}
\end{figure}

As a second demonstration of our apparatus, we injected a tilted-core vortex at the entrance, which can be described by a linear complex combination of LG\textsubscript{1,0} and LG\textsubscript{-1,0} linear modes, giving a Gaussian background and an elliptical linear core with polar and azimuthal tilt angles. With propagation in free space, the polar and azimuthal tilt angles remain constant \cite{AndersenJasmineM.2021Honv}. However, with propagation in a defocusing Kerr-type medium, calculations show that the azimuthal angle should precess \cite{ZhuChuanzhou2021Doev}. An example of the tilted-core precession is shown in Fig. \ref{fig:7}.

We chose a beam waist of $w_0=100 \mu$m to minimize the linear diffraction length $z_R$ (maximize the speed of dynamics). The linear combination of LG\textsubscript{1,0} and LG\textsubscript{-1,0} modes can be written in terms of the polar and azimuthal tilt angles, $\theta$ and $\xi$ respectively.
\begin{equation}\label{eq:9}
    \psi_0(x,y) = N \exp\bigg[-\bigg(\frac{x^2+y^2}{w_0^2} \bigg)\bigg]\frac{ax - iby}{w_0}
\end{equation}
$$a=-\cos{\xi}+i\cos{\theta}\sin{\xi}$$
$$b = -\sin{\xi}-i\cos{\theta}\cos{\xi}$$
The input polar angle was set to $\theta = 60 ^\circ$.
Intensity profiles were collected at the minimum and maximum practical propagation distances (17 mm and 117mm respectively). We took the difference of the precession angle at maximum and minimum propagation as a measure of the total precession angle at the maximum propagation distance.

To extract the precession angle from the tilted-core amplitude profiles, we chose to fit to a modified version of the nonlinear LG\textsubscript{$\pm$ 1,0} of Eq. \ref{eq:8}. We allowed the core to have some ellipticity and applied a rotation in the transverse plane.
\begin{equation}\label{eq:10}
    |\psi(x,y)_v|=N\sqrt{1-\frac{\ell^2}{x'^2+(y'\cos{\phi_{v}})^2}}
\end{equation}
$$x'=x\cos{\theta_{v}}-y\sin{\theta_{v}}$$ $$y'=x\sin{\theta_{v}}+y\cos{\theta_{v}}$$
Here, the parameter $\phi_v$ describes the ellipticity while $x'$ and $y'$ are new coordinates given by a planar rotation of angle $\theta_v$.

The background nonlinear field was also allowed some ellipticity and rotation.
\begin{equation}\label{eq:11}
    |\psi(x,y)_{bg}|=N \exp[-\bigg(\frac{\sqrt{x''^2+(y''\cos{\phi_{bg}})^2}}{w}\bigg)^g]
\end{equation}
$$x''=x\cos{\theta_{bg}}-y\sin{\theta_{bg}}$$
$$y''=x\sin{\theta_{bg}}+y\cos{\theta_{bg}}$$
Here, $\phi_{bg}$ describes the ellipticity of the background while $x''$ and $y''$ are coordinates given by a rotation of the transverse plane by angle $\theta_{bg}$. The amplitude profile used to fit the data is the product of the core and background amplitudes, $|\psi| = |\psi_v||\psi_{bg}|$.

\begin{figure}[h!]
\centering
\includegraphics[width=\linewidth]{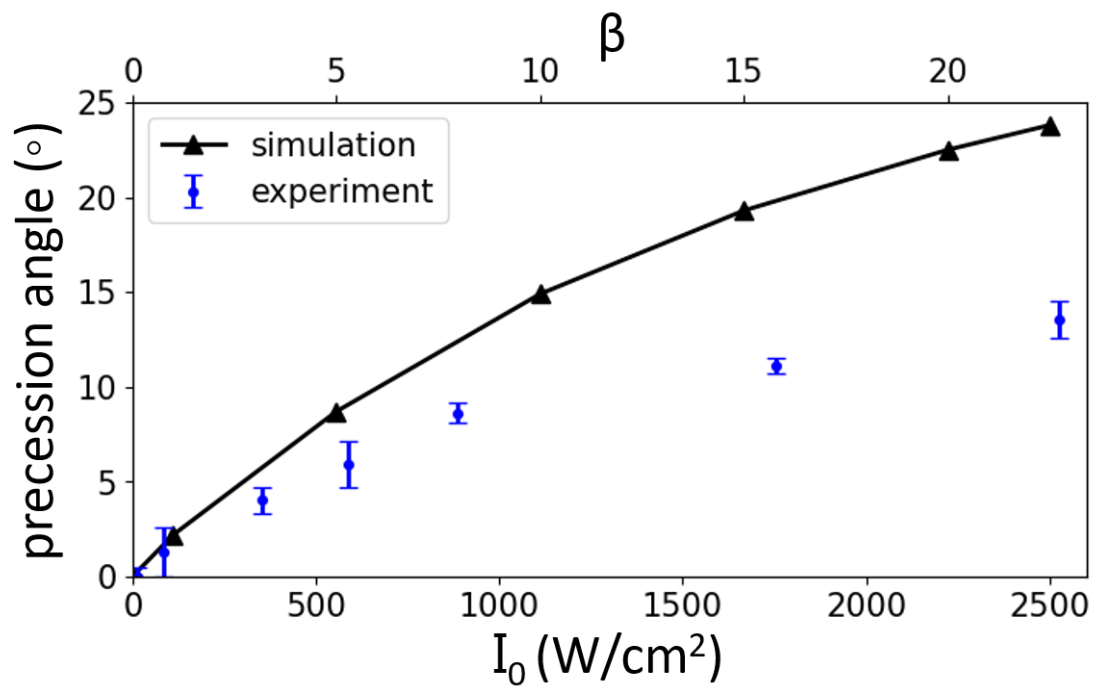}
\caption{Precession angle at a fixed distance vs $\beta$/characteristic intensity. Precession angles extracted from the numerical data shown in solid black line. Precession angles extracted from the experimental data shown in blue with error bars. The relative scaling between the $\beta$ and intensity axes is applied from the $\beta$ vs intensity ration found in the core-sharpening experiment. Error bars represent one standard deviation over 5 
measurements.}
\label{fig:8}
\end{figure}

Precession angles extracted from the experimental data are shown in Fig. \ref{fig:8} along with precession angles extracted from numerical simulation. We scaled the input intensities to $\beta$ according to the relationship found in the core-sharpening experiment, $\beta = I_0 \times 6.8\times10^{-3}$ cm\textsuperscript{2}/W.

We find that the total precession angle acquired with propagation in the optical experiment is qualitatively similar to the predictions of the GPE, with more precession for higher input intensities. The ratio of the precession angle extracted from the data divided by the precession angle from numerical simulation is approximately 60\% over the entire range of input intensities. This constant discrepancy ratio we think supports a good qualitative fit between data and model. Further, we think this discrepancy can be explained by variation in sample preparation. The relationship between input intensity and dimensionless $\beta$ was identified from the core-sharpening experiment and applied to scaling the dimensionless model for comparison with the titled-core precession data. However, the nonlinear medium was prepared for each experiment individually and could have some variation between preparations. A variation in the resulting nonlinear index $n_2$ of around 60\% seems reasonable and would explain a constant linear deviation in results across the full range of input intensities. This discrepancy in the magnitude of the precession angle may be an interesting topic for consideration in future studies.

\section{\label{sec:6}Conclusions}

We have presented a novel optical apparatus that enables the study of structured light dynamics within a nonlinear fluid medium. The instrument was used to quantify the evolution of single-charge vortex core-width and precession in a Kerr-type medium. These were found to match well with computational predictions for the mean-field behavior of quantum fluids. The strength of the optical self-interaction can be compared directly with typical values for BEC. For instance, a typical BEC ratio of system size $R$ (width of the fluid) to healing length\cite{Groszek2019} $\xi$ is $\bigl(R/\xi\bigr)_{\scriptscriptstyle \rm BEC} \sim 10-100$. In our optical system, this ratio was found to be $\bigl(R/\xi\bigr)_{\scriptscriptstyle \rm optical} \sim 4$. With further refinement of the apparatus, such as using a pulsed laser and greater demagnification of the signal beam, it should be possible to increase this ratio significantly. 

The findings of this work make it clear that optical quantum fluids support nonlinear vortex solitons. The ability to easily program essentially arbitrary initial conditions for one or more vortices, along with access to every pseudo-time slice, makes this a particularly appealing surrogate for elucidating the mean-field dynamics of quantum fluid systems that are less amenable to interrogation.

\section*{Supplemental Material}
See the supplemental material for details on non-dimensionalization of the NLSE and for further details on critical features of the experimental apparatus.

\begin{acknowledgments}
The authors acknowledge support from the W. M. Keck Foundation and the
NSF (Grant No. DMR 1553905).
\end{acknowledgments}

\section*{Author Declarations}
\subsection*{Conflict of Interests}
The authors have no conflicts of interest to disclose.
\subsection*{Author Contributions}
\textbf{Patrick Ford}: Data Curation (lead): Formal Analysis (equal): Methodology (equal): Software (equal): Validation (equal): Visualization (equal): Writing - original draft preparation (lead): Writing - review and editing (equal). \textbf{Andrew Voitiv}: Methodology (equal): Writing - review and editing (equal): \textbf{Chuanzhou Zhu}: Formal Analysis (equal): Software (equal): Validation (equal). \textbf{Mark Lusk}: Formal Analysis (equal): Funding Acquisition (equal): Project Administration (equal): Resources (equal): Supervision (equal): Validation (equal): Writing - review and editing (equal). \textbf{Mark Siemens}: Conceptualization (lead): Formal Analysis (equal): Funding Acquisition (equal): Methodology (equal): Project Administration (equal): Resources (equal): Supervision (equal): Validation (equal): Visualization (equal): Writing - review and editing (equal).

\section*{Data Availability Statement}

Data underlying all results presented are available from
the authors upon reasonable request.

\bibliography{Refs}

\pagebreak
\widetext
\begin{center}
\textbf{\large Supplemental Material \\ Measurement of nonequilibrium vortex propagation dynamics in a nonlinear medium}
\end{center}
\setcounter{equation}{0}
\setcounter{figure}{0}
\setcounter{table}{0}
\setcounter{page}{1}
\renewcommand{\theequation}{S\arabic{equation}}
\renewcommand{\thefigure}{S\arabic{figure}}
\renewcommand{\thesection}{S-\Roman{section}}
\renewcommand{\bibnumfmt}[1]{[S#1]}
\renewcommand{\citenumfont}[1]{S#1}
\renewcommand{\theequation}{S.\arabic{equation}}
\renewcommand{\thefigure}{S.\arabic{figure}}
\renewcommand{\thetable}{S.\arabic{table}}
\renewcommand{\thesection}{S.\arabic{section}}
\makeatletter
\renewcommand{\theequation}{S.\arabic{equation}}
\renewcommand{\thefigure}{S.\arabic{figure}}

\section{\label{S:1}Nondimensionalization of the nonlinear Schr\"odinger equation}
The optical nonlinear Schr\"odinger equation (NLSE) can be nondimensionalized to allow a direct comparison between the product of the characteristic intensity of the beam, $I_0$, and the material property, $n_2$, with the nonlinear coefficient, $\beta$. We look at a paraxial beam, $E(\textbf{r},z)$, as a function of transverse position, $\textbf{r} = x\hat{x} + y\hat{y}$, and axial position, $z$\, with $x$, $y$, and $z$ being orthogonal spatial dimensions.
Using the relationship $\chi^{(3)} = \frac{4}{3}n_0^2 \epsilon_0 c_0 n_2 = \frac{4}{3}n_0^2\tilde{n}_2$, we can write the NLSE of Eq. 1 in terms of the second-order refractive index:
\begin{equation}\label{eq:A1}
    i \partial_z E = \bigg( -\frac{1}{2k_0n_0}\nabla_{\perp}^2 + V - \frac{2}{3}k_0 \tilde{n}_2|E|^2\bigg)E.
\end{equation}

A key property of our experimental apparatus is the degree of nonlinearity generated. This is quantified by putting both the NLSE and the GPE in non-dimensional form and then making a direct comparison with the nonlinearity coefficient, $\beta$, of the GPE. The non-dimensional GPE , governing the dynamics of a general 2D quantum fluid $\psi'(\textbf{r}^{'},t')$, with $\textbf{r}' = x'\hat{x}+y'\hat{y}$, and a unitless time dimension $t'$ is given by
\begin{equation}\label{eq:A2}
    i \partial_t' \psi' = \bigg(-\frac{1}{2}\nabla_{\perp}^{'2} + V + \beta|\psi'|^2 \bigg)\psi'.
\end{equation}

The magnitude of $\beta$ gives the relative strength of the nonlinear interaction, and the sign of $\beta$ determines whether the interaction is attractive($\beta < 0$) or repulsive ($\beta > 0$), also referred to as self-focusing or self-defocusing, respectively.

We begin with the NLSE for optical wave propagation in the nonlinear medium, Eq. S.1, disregarding the potential term ($V(\textbf{r},z)=0$). We can re-write the NLSE by first converting propagation distance to time using $z = c_0 t/n_0$ with $c_0$ being the speed of light in free space. We then multiply through by the reduced Planck constant, $\hbar$. We can also use $k=k_0 n_0$ and $c=c_0/n_0$ to represent the wavenumber and speed of light, respectively, within the medium. Then the time derivative is
\begin{equation}\label{eq:A3}
    \partial_z = \frac{1}{c}\partial_t,
\end{equation}
and the full NLSE is
\begin{equation}\label{eq:A4}
    i \hbar \partial_t E = \bigg( -\frac{\hbar c}{2k}\nabla_{\perp}^2 - \frac{2}{3}\hbar c k\frac{\tilde{n}_2}{n_0}|E|^2\bigg)E.
\end{equation}
We recognize that we can write the electric field in terms of an amplitude and a normalized wavefunction. The input beams used in the experiment are those of the Lagurre-Gaussian modes with azimuthal index plus or minus one, and radial index zero (LG\textsubscript{$\pm$1,0}), as shown in Eq. 5. We choose to disregard the phase and look only at the amplitude of the electric field. Here the LG\textsubscript{1,0}
input beam mode of Eq. 6 is used as an example.
\begin{eqnarray}\label{eq:A5}
    |E(\textbf{r},t=0)| & = & E_0 \exp\left[-\frac{r^2}{w_0^2}\right]r \nonumber \\
    = A\frac{1}{\gamma}\exp\left[-\frac{r^2}{w_0^2}\right]r
     & =  &A|\psi(\textbf{r},t=0)| .
\end{eqnarray}

The parameter $\gamma$ above is a normalization constant ensuring $\langle  \psi | \psi \rangle = 1$. For the LG\textsubscript{1,0} input mode, $\gamma = \sqrt{\pi} w_0^2/2$. We have
\begin{equation}\label{eq:A6}
    \langle E | E \rangle = \gamma^2 E_0^2 = A^2 = \langle A\psi | A\psi \rangle .
\end{equation}
Returning to the dynamics and substituting for the electric field, we have
\begin{equation}\label{eq:A7}
    i \hbar \partial_t \psi = \bigg( -\frac{\hbar c}{2k}\nabla_{\perp}^2 - \frac{2}{3}\hbar c k\frac{\tilde{n}_2}{n_0}A^2|\psi|^2\bigg) \psi.
\end{equation}

A choice of observable-parameter units allows nondimensionalization.
\begin{eqnarray}\label{eq:A8}
    & &[d] \sim \frac{1}{k} \nonumber \\
    & &[t] \sim \frac{1}{ck} \\
    & &[E] \sim \hbar c k \nonumber .
\end{eqnarray}
The above choice of units give the following conversions:
\begin{eqnarray}\label{eq:A9}
    \partial_t &=& [t]^{-1}\partial_{t'} = ck \partial_{t'} \nonumber\\
    \nabla^2 &=& [d]^{-2} \nabla^{'2} = k^2\nabla^{'2} \\
    |\psi|^2 &=& [d]^{-2}|\psi'|^2 = k^2|\psi'|^2 .\nonumber
\end{eqnarray}

Substituting these relationships into the dynamics gives
\begin{equation}\label{eq:A10}
    i  \partial_{t'}\psi' = \bigg( -\frac{1}{2}\nabla_{\perp}^{'2} - \frac{2}{3}\frac{\tilde{n}_2}{n_0}A^2 k^2 |\psi'|^2\bigg)\psi'.
\end{equation}
We can relate the quantity $A^2$ to the input beam power using
\begin{eqnarray}\label{eq:A11}
    P &=& \int_{All}Ida = \frac{\epsilon_0 c_0 n_0}{2} \langle E| E \rangle \nonumber \\
    &=& \frac{\epsilon_0 c_0 n_0}{2} \langle A\psi| A\psi \rangle
\end{eqnarray}
to write the amplitude in terms of the beam's characteristic intensity, $A^2 = 2\gamma^2 I_0/\epsilon_0 c_0 n_0$. Applying this relationship to the dynamics along with $\tilde{n}_2 = \epsilon_0 c_0 n_2$ yields
\begin{equation}\label{eq:A12}
    i  \partial_{t'}\psi' = \bigg( -\frac{1}{2}\nabla_{\perp}^{'2} - \frac{4}{3}k_0^2 n_2\gamma^2 I_0 |\psi'|^2\bigg)\psi'
\end{equation}

We began with the NLSE for classical paraxial propagation of coherent light in a $\chi^{(3)}$ medium, investigated the relationship between the electric field and a normalized wavefunction, and chose appropriate units to arrive at a dimensionless form for propagation dynamics. It is evident that the non-dimensional NLSE is directly analogous to the GPE. Further, by inspection we can identify an equivalence between the $\beta$ in the GPE and the nonlinear coefficient in the non-dimensional NLSE;
\begin{equation}\label{eq:A13}
    \beta = -\frac{4}{3}k_0^2 n_2 \gamma^2 I_0
\end{equation}

\section{\label{S:2}Nonlinear fluid tank}
The fluid tank is constructed from off-the shelf parts to a large degree. A custom window aperture and dynamic piston seal aperture were fabricated from drawings. The imaging tube was fabricated by modifying a 1-inch diameter optics tube.

The tank itself is a cast aluminum project box with a close-fitting lid. The interior of the project box container is 7 inches long, 4.5 inches wide, and 3.5 inches deep. Holes were drilled in the ends of the box opposite each other in the long direction. The custom-fabricated window and tube apertures are epoxied into these holes to allow fitting the input window and telescoping imaging tube to the container.
\begin{figure}[h!]
\centering
\includegraphics[width=\linewidth]{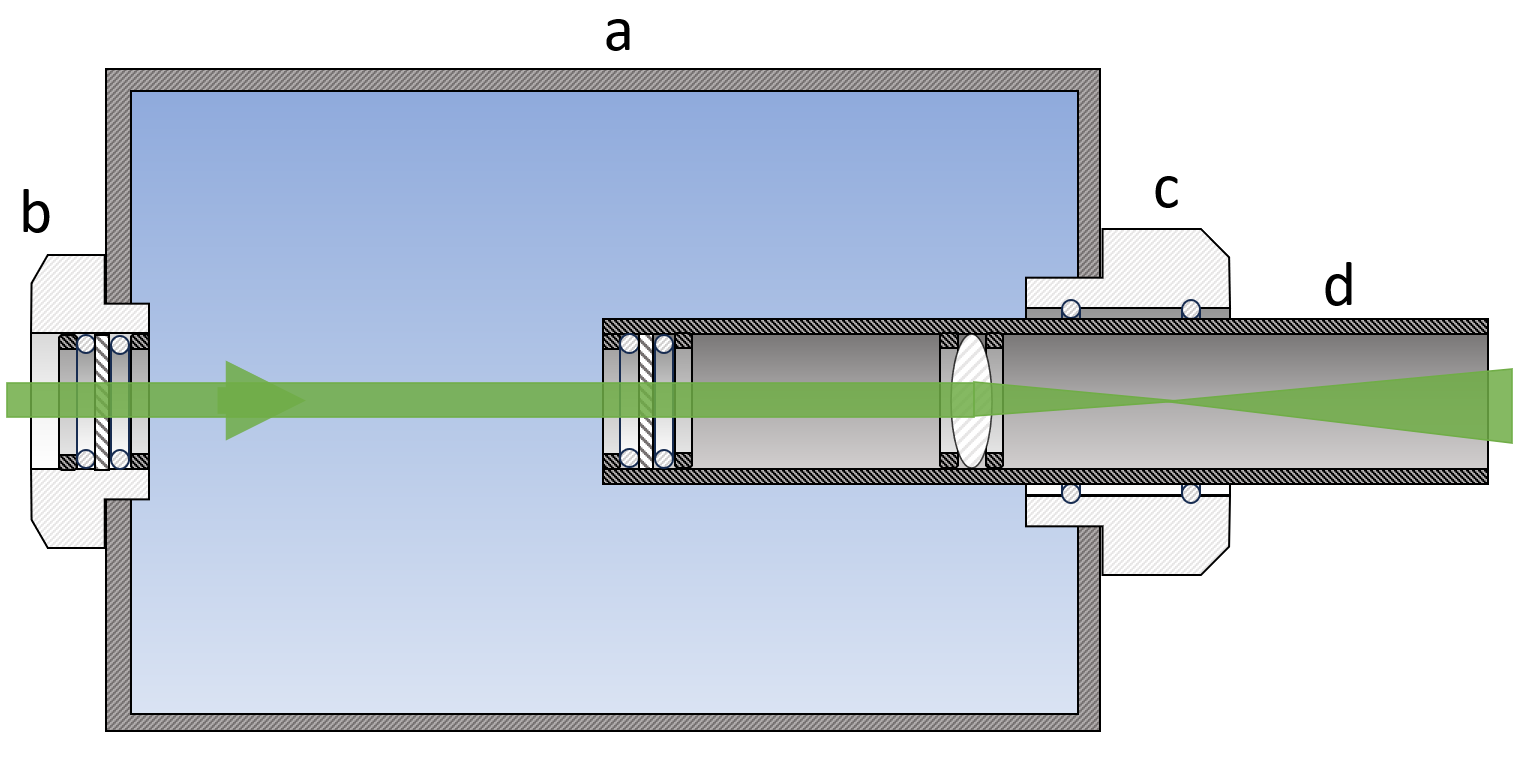}
\caption{Fluid tank detail. The tank (a) itself holds the GO-doped ethanol solution. The window aperture (b) secures a window sandwiched between o-rings and retaining rings. A tube aperture (c) provides a dynamic piston seal against the imaging tube (d) using o-rings and designed tolerances. The imaging tube holds a window on the tank side sandwiched between o-rings and retaining rings. The lens of a single-lens imaging system is held within the tube by retaining rings. The tube can translate in the propagation direction which provides a variable propagation distance for the beam within the nonlinear fluid. The magnified beam is imaged onto a camera sensor for data capture.}
\label{fig:9}
\end{figure}

The window aperture was specified to fit a standard 1-inch diameter optical window (Thorlabs WG11010-A). The aperture is threaded to SM1 specifications. PTFE o-rings are used to seal the window aperture. One o-ring (McMaster Carr 9559K63) is placed on each side of the window, and an SM1 retaining ring (Thorlabs SM1RR) is used to compress the o-rings and window in place.

The imaging tube aperture uses dynamic piston seal tolerances selected for the outer diameter of the imaging tube. The aperture includes two inner-circumference grooves to retain the appropriate PTFE o-rings (McMaster Carr 955K158) and provide a water-tight dynamic seal against the outside of the imaging tube. Dielectric grease is placed between the o-rings to aid in sealing and provide some lubrication for the translating imaging tube.

The imaging tube is fabricated from a 1-inch diameter 6-inch long SM1 lens tube extension. The interior is modified to include SM1 threads along the entire length of the inside of the tube. At one end of the tube, a window is secured using a retaining ring/o-ring/window/o-ring/retaining ring sandwich. Within the tube, a $100$ mm lens (Thorlabs LA1509-A) is placed to provide a single-lens imaging system. The inner lens is held between two SM1 retaining rings.

The tank is mounted to the optical bench with the imaging tube sticking out toward the camera. The window end of the tube is situated inside the tank. The outside end of the tube is attached to a linear motion computer-controlled stage with 500 mm of travel. The image attenuation, interference elements, and camera were also mounted to the motion stage such that the tube, imaging system, and camera move together. The result of moving the stage is a change in the propagation distance within the nonlinear medium, while still having a fixed imaging system in free air from the output window to the camera.

The reference beam propagation axis is aligned parallel to the signal beam path, and a mirror re-directs the reference beam onto beam splitter for interference as shown in Fig. 2. Careful alignment allows producing interferograms with variable nonlinear propagation.

\nocite{*}

\end{document}